\documentclass[preprint,aps,showpacs]{revtex4-1}
\usepackage{amsmath,amssymb,amsfonts,dcolumn,color,graphicx,graphics,latexsym,placeins,epsfig,tikz}
\usetikzlibrary{arrows,shapes}
\usepackage{subfigure,rotating,hyperref,bm,array}

\newcommand{\be}{\begin{equation}}
\newcommand{\ee}{\end{equation}}
\newcommand{\ba}{\begin{eqnarray}}
\newcommand{\ea}{\end{eqnarray}}

\begin{document}

\title{\Large \bf Gravitational lensing by scalar-tensor wormholes 
and the energy conditions}

\author{Rajibul Shaikh}
\email{rajibulshaikh@cts.iitkgp.ernet.in}
\thanks{\\Present address: Department of Astronomy $\&$ Astrophysics, Tata Institute of Fundamental Research, Homi Bhabha Road, Mumbai 400005, India.} \email{rajibul.shaikh@tifr.res.in}
\author{Sayan Kar}
\email{sayan@phy.iitkgp.ernet.in}
\affiliation{${}^{*}$ Centre for Theoretical Studies, Indian Institute of Technology Kharagpur, Kharagpur, 721 302, India.}
\affiliation{${}^{\dagger}$ Department of Physics {\it and} Centre for Theoretical Studies \\ Indian Institute of Technology Kharagpur, Kharagpur, 721 302, India.}

\begin{abstract}
We study gravitational lensing by a class of zero Ricci scalar wormholes 
which arise as solutions in a scalar-tensor theory of gravity.  
An attempt is made to find a possible link between 
lensing features, stable/unstable photon orbits 
and the energy conditions on the matter required 
to support these spacetimes. Our analysis shows (for this class of wormholes) 
that light rays always exhibit
a positive deflection if the energy conditions are satisfied 
(nonexotic matter content). In contrast, if the energy conditions are 
violated (exotic matter), the net deflection of a light ray
may be positive, negative or even zero, 
depending on values of the metric and impact parameters. 
This prompts us to introduce a surface defined by a 
turning point value at which the net deflection 
of a light ray is equal to zero, even though we
have a curved spacetime geometry. 
We argue that the existence of such a surface may be linked to 
exotic/energy condition violating matter. Wormholes in modified gravity
with matter satisfying the energy conditions do not seem to have such a
zero deflection surface. 
Finally, we study strong gravitational lensing briefly and also look into 
the formation of Einstein and relativistic Einstein rings.
We conclude with some estimates on the wormhole mass, throat-radius and 
the detectability of the Einstein rings.
  
\end{abstract}


\maketitle

\section{Introduction}
Gravitational lensing is one of the observational probes of a given 
spacetime geometry and the gravitational field it represents. 
Theoretical prediction of light bending and its subsequent observational 
verification \cite{Dyson,Eddington} is celebrated as one of the successful 
tests of general relativity (GR) in the weak field limit. In modern astronomy
and cosmology, gravitational lensing is widely used as an important tool 
for probing extrasolar planets, highly redshifted galaxies, quasars, 
supermassive black holes, dark matter candidates, primordial gravitational 
wave signatures, etc. \cite{lensing_application1,lensing_application2,lensing_application3}. It is also used to test 
the viability of different alternative theories of gravity \cite{GTT1}.

After the observational confirmation of the bending of light, 
interest in studying gravitational lensing has grown immensely 
in the last few decades. 
Strong deflection of light rays by the simplest black hole, the 
Schwarzschild, was first studied in \cite{darwin}. 
An exact analytic expression for the deflection angle by the Schwarzschild 
black hole was obtained in \cite{ohanian}. Subsequently, gravitational lensing by the Schwarzschild black hole in the strong field limit and various aspects (eg. formation of relativistic images, their magnification and critical curves) have been studied in \cite{newman,virbhadra_schwarzschild1,bozza_schwarzschild1,virbhadra_schwarzschild2} in a systematic way. 
Bozza developed an analytic framework to study gravitational lensing for a 
completely generic, spherically symmetric, static spacetime in the strong 
field limit and applied it to some existing spacetimes \cite{bozza_general1,bozza_general2}. The study of gravitational lensing is not only limited to 
the Schwarzschild black hole. Lensing by other black holes such as 
Reissner-Nordstr$\ddot{\text{o}}$m \cite{reissner1,reissner2,reissner3,PhysRevD.95.064034}, Kerr \cite{kerr1,kerr2,kerr3}, Kiselev \cite{kiselev,kiselev_charged}, global monopole \cite{global_monopole1,global_monopole2,global_monopole3}, Einstein-Born-Infeld \cite{einstein_born_infeld}, Eddington-Born-Infeld \cite{eddington_born_infeld1,eddington_born_infeld2,eddington_born_infeld3}, scalar-tensor \cite{scalar_tensor1,scalar_tensor2}, braneworld \cite{braneworld1,braneworld2,braneworld3,sumanta_lensing}, dilaton \cite{dilaton1,dilaton2,dilaton3,dilaton4}, phantom \cite{phantom1,phantom2}, regular \cite{regular1,regular2,regular3,regular4}, Kaluza-Klein \cite{kaluza_klein1,kaluza_klein2,kaluza_klein3,kaluza_klein4}, Horava-Lifshitz \cite{horava_lifshitz}, Myers-Perry \cite{myers_perry} and Galileon \cite{galileon} black holes, have been 
studied. Gravitational lensing by naked singularities has been analyzed
to see whether one can distinguish between a black hole and a naked singularity \cite{naked1,naked2,naked3,naked4,naked5}. Such studies show that, in some 
cases, lensing by a naked singularity is qualitatively different from that 
by a black hole. Gravitational lensing \cite{WL1,WL2,WL3,WL4,WL5,WL6,WL7,WL8,WL9,WL10,PhysRevD.95.064035,PhysRevD.95.084021} and microlensing \cite{WL11,WL12,WL13} by wormholes has been a topic of growing interest in the recent past too. 
It has been shown that gravitational lensing can be used as a 
distinguishing probe between black holes and wormholes 
\cite{WL12,harada1,harada2}.

However, the above-mentioned works are limited only to lensing features. 
Any possible link between lensing and matter content of a given spacetime 
geometry, has not been explored much. It may be noted that though 
gravitational lensing can be understood geometrically, 
the field equations of a given theory of gravity relates geometry to 
matter stress energy. Thus, gravitational lensing properties of a given 
spacetime geometry should, in principle, be related to the matter content 
of the spacetime. Attempts have been made to relate lensing with matter 
stress energy within the framework of GR \citep{lensing_matter1,lensing_matter2,lensing_matter3}. In our work here, we explore the relation between lensing 
and matter content through a specific example of a geometry in 
on-brane scalar-tensor gravity. In particular, we study gravitational lensing 
by $R=0$ on-brane scalar-tensor wormholes obtained in \cite{RS1}. It is 
well-known that wormholes in GR, violate the energy conditions. However, the 
modified or alternative gravity wormholes do not necessarily violate the 
energy conditions (see \cite{nonexotic_wormholes} for some recent examples 
and references therein). Depending on the parameter values, the $R=0$ 
wormholes in our study, may or may not violate the energy conditions. 
Therefore, it is of interest to explore whether there is any quantitative 
or qualitative difference between lensing by a wormhole satisfying the energy 
conditions and that for wormholes violating them.

\noindent Our paper is organized as follows. In Sec. \ref{sec:energy conditions}, we briefly recall the $R=0$ scalar-tensor wormholes and the corresponding energy conditions. In Sec. \ref{sec:photon orbit}, we study the existence of stable and unstable photon orbits and its relation with the energy conditions. Sections \ref{sec:lensing} and \ref{sec:strong deflection} deal with gravitational lensing by the general $R=0$ wormholes. In Secs. \ref{sec:Einstein rings} and \ref{sec:detectability}, we study the formation of Einstein and relativistic Einstein rings and their detectability. Finally, we conclude in Sec. \ref{sec:conclusion}. 

\section{$\text{R=0}$ wormholes and the energy conditions}
\label{sec:energy conditions}
The $R=0$ spacetime geometry obtained in \cite{RS1} is given by
\begin{equation}
ds^2=-f^2(r)dt^2+\frac{dr^2}{1-\frac{b(r)}{r}}+r^2\left(d\theta^2+\sin^2\theta d\phi^2\right),
\label{eq:metric1}
\end{equation}
where $b(r)=2m+\frac{\beta}{r}$ and
\begin{equation}
f(r)=\frac{1}{(1+\eta)}\left[1+\frac{\beta}{m^2}\frac{m}{r}+\eta\sqrt{1-\frac{2m}{r}-\frac{\beta}{m^2}\frac{m^2}{r^2}}\right].
\end{equation}
The $\beta=0$ version of this geometry has been discussed in {\cite{dadhich}} 
and, more recently, in \cite{skslsg}.

\begin{figure}[ht]
\centering
\includegraphics[scale=1.0]{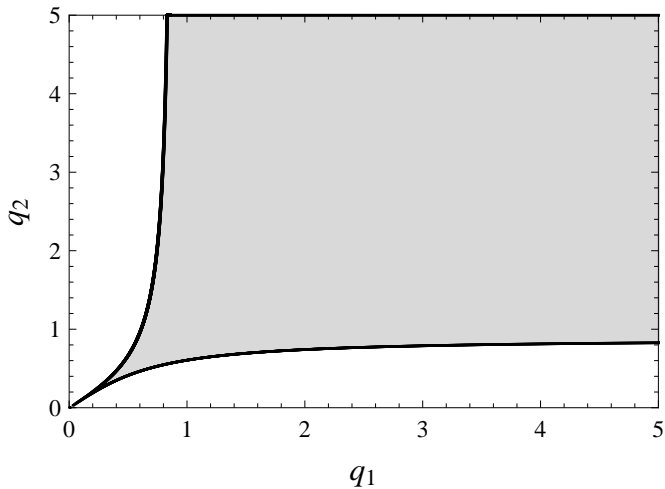}
\caption{Parameter regions (shaded) for which the energy conditions are satisfied.}
\label{fig:region0}
\end{figure}

The $R=0$ equation which will be useful in our subsequent analysis is given by
\begin{equation}
\left(1-\frac{b}{r}\right)f''(r)+\frac{4r-3b-b'r}{2r^2}f'(r)-\frac{b'}{r^2}f(r)=0.
\label{eq:R0}
\end{equation}
In \cite{RS1}, the weak energy condition was studied by writing down the solution in isotropic coordinates. The transformation from the Schwarzschild radial coordinate $r$ to the isotropic coordinate $\bar{r}$ is given as
\begin{equation}
r=\left (1+\frac{m}{\bar{r}}+\frac{m^2+\beta}{4\bar{r}^2}\right)\bar{r}.
\end{equation}
The line element in isotropic coordinates, becomes
\begin{equation}
ds^2 = -\frac{h^2(\bar{r})}{U^2(\bar{r})}dt^2 + U^2(\bar{r})\left (d\bar{r}^2+\bar{r}^2d\Omega_2^2\right ),
\end{equation}
where
\begin{equation}
h(x)= \frac{1-\eta}{1+\eta} \left (q_1+x\right )\left (q_2+x\right ),
\end{equation}
\begin{equation}
U(x)=1 + x^2+2\mu x,
\end{equation}
and 
\begin{equation}
x=\frac{\sqrt{m^2+\beta}}{2\bar{r}},\hspace{0.3in} \mu = 
\frac{m}{\sqrt{m^2+\beta}}.
\end{equation}
The constants $q_1$ and $q_2$ are given as
\begin{equation}
q_1= \frac{\mu(1+\eta)}{1-\sqrt{\mu^2\eta^2+1-\mu^2}},\hspace{0.3in}
q_2= \frac{\mu(1+\eta)}{1+\sqrt{\mu^2\eta^2+1-\mu^2}}.
\end{equation}
It is easy to show that $\eta$ and $\mu$ are related to $q_1$ and $q_2$ 
as follows:
\begin{equation}
\mu=\frac{q_1 q_2+1}{q_1+q_2},\hspace{0.3in}
\eta=\frac{q_1 q_2-1}{q_1q_2+1}.
\label{eq:mu_eta}
\end{equation}
In \cite{RS1}, we have shown that, in order to satisfy the weak energy condition, we must have, in the limit $x\to 0$ (i.e., $r\to\infty$),
\begin{equation}
\gamma^2= \frac{(q_1-q_2)^2(q_1+q_2)}{4(q_1+q_2)(\log\frac{q_1}{q_2})^2 -8(\log\frac{q_1}{q_2}) (q_1-q_2)},
\label{eq:gamma1}
\end{equation}
where $\gamma$ is an integration constant related to the scalar field given by
\begin{equation}
\xi (x)=\frac{2\gamma}{q_1-q_2}\log\left\vert\frac{q_1+x}{q_2+x}\right\vert.
\end{equation}
Using Eqn. (\ref{eq:gamma1}), it can be shown that, in the limit $x\to 0$ (i.e., for large $r$), the energy density $\rho$, radial pressure $\tau$ and the tangential pressure $p$ (see \cite{RS1} for the expressions of $\rho$, $\rho+\tau$ and $\rho+p$) behave as
\begin{equation}
\rho=\frac{Q^2}{r^4}+\mathcal{O}\left(\frac{1}{r^5}\right), \hspace{0.3cm} \tau=-\frac{Q^2}{r^4}+\mathcal{O}\left(\frac{1}{r^5}\right), \hspace{0.3cm} p=\frac{Q^2}{r^4}+\mathcal{O}\left(\frac{1}{r^5}\right),
\nonumber
\end{equation}
where
\begin{equation}
Q^2=\frac{16\mu^2 l}{m^2\kappa^2}\left[(\mu^2-1)(\xi_0^2-1)+\frac{\gamma^2}{q_1^2 q_2^2}-\frac{\gamma\xi_0}{q_1^2 q_2^2}(q_1+q_2-2\mu q_1 q_2)\right],
\nonumber
\end{equation}
and $\xi_0=\xi(x=0)$. Therefore, asymptotically, matter energy-momentum resembles that due to a static electric field in Maxwell electrodynamics 
with $Q$ playing the role of electric charge. In \cite{RS1}, we have also shown that, in the limit $x\to 0$, the positivity of the leading-order term in $\rho$ (i.e., $Q^2$ in the present work) ensures the satisfaction of the weak energy condition. However, there we set $Q^2$ to zero and obtained another expression for $\gamma^2$ (Eqn. (28) of \cite{RS1}). By comparing the two expressions for $\gamma^2$, we obtained a relation between $q_1$ and $q_2$. However, in the present work, we relax this condition and do not set $Q^2$ to zero identically. We choose $q_1$ and $q_2$ independently in such a way that $Q^2\geq 0$. This gives
\begin{equation}
\frac{2\gamma^2}{(q_1^2-q_2^2)}\left[4(\mu^2-1)q_1^2q_2^2-(q_1+q_2)^2+2q_1q_2(1+q_1q_2)\right]\log\left\vert\frac{q_1}{q_2}\right\vert+\gamma^2\geq 0,
\end{equation}
which allows us to explore a wider range of the parameter space for which 
the weak energy condition holds. Note that the traceless condition $-\rho+\tau+2p=0$ implies $\rho+\tau+2p=2\rho$. Hence, the satisfaction of the weak energy condition ensures the satisfaction of the strong and hence, all other energy 
conditions. The shaded region in Fig. \ref{fig:region0} shows the parameter 
values for which the energy conditions are satisfied.

\section{Stable and unstable photon orbits and the energy conditions}
\label{sec:photon orbit}
\begin{figure}[ht]
\centering
\includegraphics[scale=1.0]{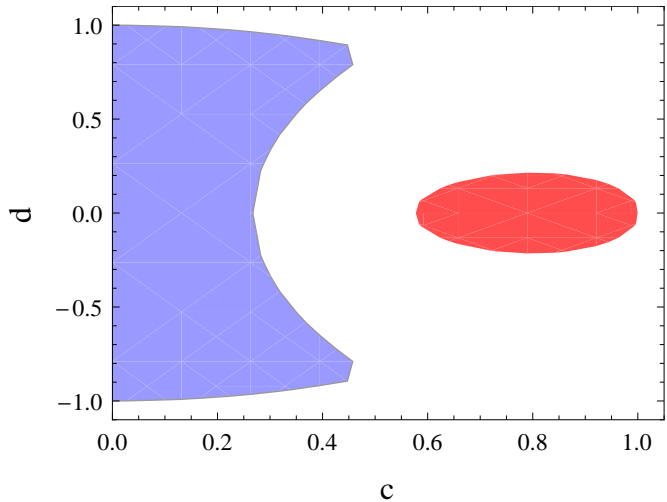}
\caption{Parameter regions showing the presence of stable photon orbit, i.e., $2\leq \mu< \infty$ and $-1<\eta\leq 0$ (blue) and satisfaction of the energy conditions (red). The absence of an overlap between blue and red regions indicates that the energy conditions have to be violated for parameter values required to have a stable photon orbit.}
\label{fig:region2}
\end{figure}

The first integral obtained from the geodesic equations for a photon moving in the 
equatorial ($\theta=\frac{\pi}{2}$) plane of the spacetime in Eqn. (\ref{eq:metric1}) are given by
\begin{equation}
\dot{t}=\frac{E}{f^2(r)}, \hspace{0.3cm} \dot{\phi}=\frac{L}{r^2}, \hspace{0.3cm} \frac{f^2(r)\dot{r}^2}{1-\frac{b(r)}{r}}+V(r)=E^2,
\end{equation}
where the energy $E$ and the angular momentum $L$ are constants of motion. An overdot represents differentiation with respect to the parameter
$\lambda$ parametrizing the geodesics. 
The effective potential $V(r)$ is given by
\begin{equation}
V(r)=\frac{L^2}{r^2}f^2(r)
\end{equation}
Circular photon orbits correspond to $\dot{r}=0$ (i.e., $V=E^2$) and $V'(r)=0$ (i.e., extrema of the effective potential). Stable and unstable photon orbits correspond to the minimum ($V''(r)>0$) and maximum ($V''(r)<0$) of the effective potential, respectively. Defining $z=\frac{r_0}{r}=\frac{m+\sqrt{m^2+\beta}}{r}=\frac{m(1+\mu)}{\mu r}$, the effective potential can be rewritten as
\begin{equation}
V(z)=\frac{\mu^2 l^2 z^2}{(1+\mu)^2}f^2(z),
\end{equation}
where
\begin{equation}
f(z)=\frac{1}{(1+\eta)}\left(1+\frac{1-\mu}{\mu}z+\eta\sqrt{1-\frac{2\mu}{1+\mu}z-\frac{1-\mu}{1+\mu}z^2} \right),
\end{equation}
$l=\frac{L}{m}$ and $r_0$ is the throat radius.  The stable and unstable photon orbits still correspond to $\frac{d^2V}{dz^2}>0$ and $\frac{d^2V}{dz^2}<0$, respectively. The extrema $z_{ex}$ of the potential are given by
\begin{equation}
\frac{1}{f(z_{ex})}\frac{df}{dz}\Big\vert_{z_{ex}}+\frac{1}{z_{ex}}=0
\label{eq:extremum1}
\end{equation}
which becomes
\begin{equation}
\eta \frac{\left(1-\frac{3\mu}{1+\mu}z_{ex}-2\frac{1-\mu}{1+\mu}z_{ex}^2\right)}{\sqrt{1-\frac{2\mu}{1+\mu}z_{ex}-\frac{1-\mu}{1+\mu}z_{ex}^2}}+\left(1+2\frac{1-\mu}{\mu}z_{ex}\right)=0.
\label{eq:extremum2}
\end{equation}
Using Eqns. (\ref{eq:R0}) and (\ref{eq:extremum1}), it can be shown that \cite{RS1}
\begin{equation}
\frac{d^2V}{dz^2}\Big\vert_{z_{ex}}=-\frac{4l^2\mu^3}{(1+\mu)^3}\frac{f^2(z_{ex})}{1-\frac{2\mu}{1+\mu}z_{ex}-\frac{1-\mu}{1+\mu}z_{ex}^2}\left(\frac{1+\mu}{\mu}-\frac{3}{2}z_{ex}\right).
\end{equation}
Therefore, minima ($\frac{d^2V}{dz^2}>0$) exist if $z_{ex}$ are real and $z_{ex}>\frac{2(1+\mu)}{3\mu}$. Note that we must have $0\leq z_{ex}\leq 1$. This is possible when $\mu>2$. Let us now put $z_{ex}=\frac{2(1+\mu)}{3\mu}$ in 
Eqn. (\ref{eq:extremum2}). We obtain
\begin{equation}
\eta=\frac{1-\frac{4}{3}\frac{\mu^2-1}{\mu^2}}{1-\frac{8}{9}\frac{\mu^2-1}{\mu^2}}\sqrt{\frac{4}{9}\left(\frac{\mu^2-1}{\mu^2}\right)-\frac{1}{3}}.
\label{eq:minima_condition}
\end{equation}
Note that $\eta=0$ at $\mu=2$ and $\eta\to-1$ as $\mu\to \infty$. Therefore, the ranges to have stable photon orbits, become $2\leq \mu<\infty$ and $-1<\eta\leq 0$. Using Eqn. (\ref{eq:minima_condition}), one can show that $(\mu^2\eta^2+1-\mu^2)$ is always negative ($\leq -3$ for $\mu\geq 2$) by plotting it as a function of $\mu$. This implies that $q_1$ and $q_2$ become complex conjugates of each other. Therefore, stable photon orbit does not exist for real $q_1$ and $q_2$ satisfying the energy conditions. It may exist for imaginary $q_1$ and $q_2$. Note that, for complex conjugate $q_1$ and $q_2$, the metric functions, the energy density and pressures are real since $q_1 q_2$ and $(q_1+q_2)$ are real. It can also be shown that $\xi(x)$ is real. To verify this, let $q_1=c+id$ and $q_2=c-id$, where $c$ and $d$ are real. In terms of $c$ and $d$, $\mu$ and $\eta$ become
\begin{equation}
\mu=\frac{c^2+d^2+1}{2c}, \hspace{0.5cm} \eta=\frac{c^2+d^2-1}{c^2+d^2+1}
\end{equation}
Note that we must have $c>0$ since $\mu>0$. We express $(q_1+x)$ and $(q_2+x)$ as
\begin{equation}
(q_1+x) = \left\{
  \begin{array}{lr}
   \sqrt{(c+x)^2+d^2}e^{i\theta(x)} & : c>0, d>0\\
   \sqrt{(c+x)^2+d^2}e^{i(2\pi-\theta(x))} & : c>0, d<0
  \end{array}
\right. ,
\end{equation}
\begin{equation}
(q_2+x) = \left\{
  \begin{array}{lr}
   \sqrt{(c+x)^2+d^2}e^{i(2\pi-\theta(x))} & : c>0, d>0\\
   \sqrt{(c+x)^2+d^2}e^{i\theta(x)} & : c>0, d<0
  \end{array}
\right. ,
\end{equation}
where $\theta(x)=\tan^{-1} \left(\frac{|d|}{|c+x|}\right)$. Therefore, $\xi(x)$ becomes
\begin{equation}
\xi(x) = \left\{
  \begin{array}{lr}
   \frac{2\gamma(\theta(x)-\pi)}{d} & : c>0, d>0\\
   \frac{2\gamma(\pi-\theta(x))}{d} & : c>0, d<0
  \end{array}
\right. ,
\end{equation}
which is real. To make $\xi(x)$ positive, we must have $\gamma<0$ since $0\leq \theta(x)\leq \frac{\pi}{2}$. This is achieved by taking the negative root of 
Eqn. (\ref{eq:gamma1}), which becomes
\begin{equation}
\gamma^2 = \left\{
  \begin{array}{lr}
   \frac{1}{2}\frac{d^2}{(\theta_0-\pi)^2-\frac{d}{c}(\theta_0-\pi)} & : c>0, d>0\\
   \frac{1}{2}\frac{d^2}{(\pi-\theta_0)^2-\frac{d}{c}(\pi-\theta_0)} & : c>0, d<0
  \end{array}
\right. ,
\end{equation}
where $\theta_0=\tan^{-1}\left(\frac{|d|}{|c|}\right)$. In the limit $x\to 0$, one can also express $Q^2$ in terms of $c$ and $d$. Figure \ref{fig:region2} shows the parameter region where $2\leq \mu< \infty$ and $-1<\eta\leq 0$ and the region where the energy conditions are satisfied. The absence of an overlap region indicates that the energy conditions have to be violated for the parameter values required to have a stable photon orbit. Therefore, no stable photon orbit exists for the parameter values satisfying the energy conditions. However, as shown in \cite{RS1}, unstable photon orbits (photon spheres) exist for 
parameter values satisfying the energy conditions.

\section{Gravitational lensing by the $R=0$ wormholes and the energy conditions}
\label{sec:lensing}
\begin{figure}[ht]
\centering
\includegraphics[scale=1.0]{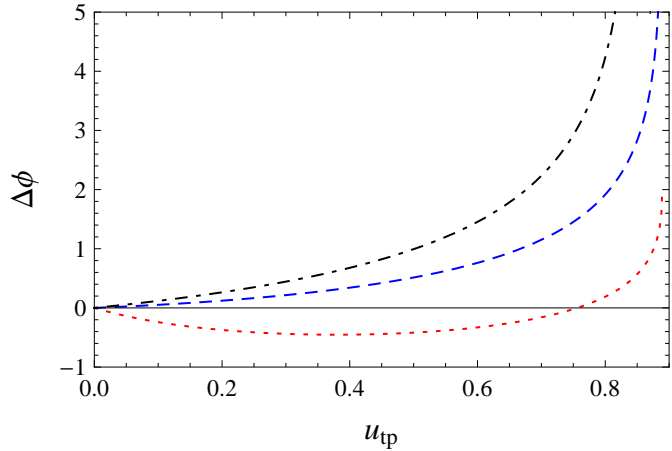}
\caption{Plots of deflection angle (obtained by numerical integration) as a function of the inverse turning point $u_{tp}$($=\frac{1}{r_{tp}}$) for $\eta=0.7$ (black dot-dashed), $\eta=0.0$ (blue dashed) and $\eta=-0.7$ (red dotted). Here, we have taken $\mu=0.8$ and $m=0.5$. For all these $\mu$ and $\eta$ values, $q_1$ and $q_2$ are real and positive.}
\label{fig:sd}
\end{figure}

\begin{figure}[ht]
\centering
\includegraphics[scale=0.98]{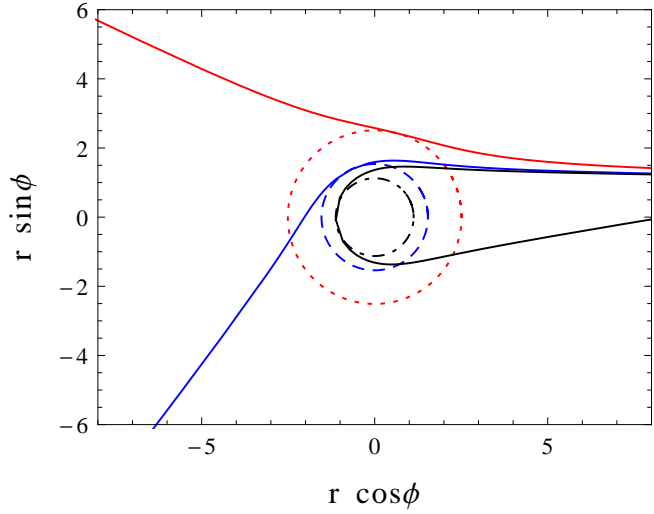}
\caption{The equatorial trajectories of light rays for $\eta=-0.7$ (red), $\eta=0.0$ (blue) and $\eta=0.7$ (black). Here, we have taken $\mu=0.8$, $m=0.5$ and $b=1.3$. The circles show the corresponding turning points. The light source is at $r=100$ and $\phi=0$.}
\label{fig:light_ray0}
\end{figure}

\begin{figure}[ht]
\centering
\includegraphics[scale=1.0]{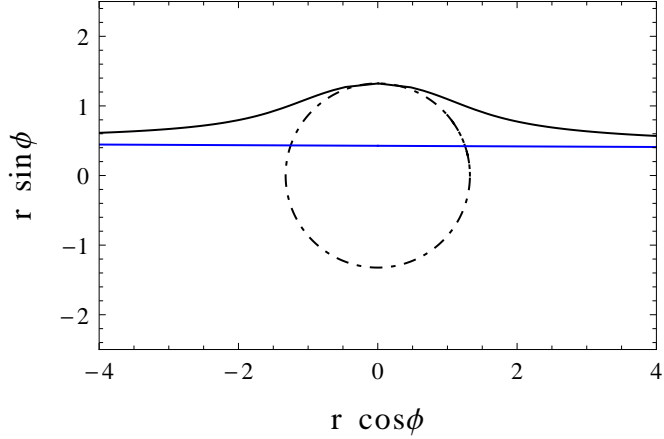}
\caption{The equatorial trajectory of a light ray undergoing zero net deflection. The blue line shows an undeflected light ray. Here, we have taken $\mu=0.8$, $\eta=-0.7$, $m=0.5$ and $b=0.427$. The circle shows the corresponding turning point. The light source is at $r=100$ and $\phi=0$.}
\label{fig:light_ray00}
\end{figure}

Let us now see how the $R=0$ wormholes affect light rays. We begin with a general spherically symmetric, static spacetime given by
\begin{equation}
ds^2=-A(r)dt^2+B(r)dr^2+C(r)d\Omega^2.
\label{eq:arbitrary_metric}
\end{equation}
The first integrals obtained from the geodesic equations for a photon moving in the equatorial plane ($\theta=\frac{\pi}{2}$ plane) are given by
\begin{equation}
\dot{t}=\frac{E}{A(r)}, \hspace{0.3cm} \dot{\phi}=\frac{L}{C(r)}, \hspace{0.3cm} \dot{r}=\pm\frac{1}{\sqrt{B(r)}} \sqrt{\frac{E^2}{A(r)}-\frac{L^2}{C(r)}}.
\end{equation}
Therefore, we obtain
\begin{equation}
\frac{d\phi}{dr}=\pm \frac{\sqrt{B(r)}}{C(r)\sqrt{\frac{1}{A(r)b^2}-\frac{1}{C(r)}}},
\end{equation}
where $b=\frac{L}{E}$ is the impact parameter. At the turning point $r_{tp}$, $\frac{dr}{d\phi}\big|_{r_{tp}}=0$. This gives the following relationship between the impact parameter $b$ and the turning point $r_{tp}$: 
\begin{equation}
b=\sqrt{\frac{C(r_{tp})}{A(r_{tp})}}.
\end{equation}
Therefore, the exact expression for deflection angle can be written as \cite{wald}
\begin{equation}
\Delta\phi=2I(r_{tp})-\pi,
\end{equation}
where
\begin{equation}
I(r_{tp})=\vert\phi(\infty)-\phi(r_{tp})\vert
\end{equation}
and
\begin{equation}
\phi(r)=\int^r \frac{\sqrt{B}}{\sqrt{C}\sqrt{\frac{C(r)A(r_{tp})}{C(r_{tp}) A(r)}-1}}dr.
\end{equation}
It is difficult to perform the above integration for the general $R=0$ metric. The case $\mu=1$ and $\eta=0$ gives the ultrastatic Schwarzschild wormhole. In 
this case, we put $\sin t=\sqrt{\frac{1-\frac{2m}{r}}{1+\frac{2m}{r_{rtp}}}}$ in the integral. We obtain
\begin{eqnarray}
\phi(r)&=& \frac{2i}{\sqrt{1-\frac{2m}{r_{tp}}}}\int^r \frac{1}{\sqrt{1+\frac{1+\frac{2m}{r_{tp}}}{1-\frac{2m}{r_{tp}}}\sin^2t}}dt  \\
&=&\frac{2i}{\sqrt{1-\frac{2m}{r_{tp}}}}{\rm EllipticF}\left[\sin^{-1}\sqrt{\frac{1-\frac{2m}{r}}{1+\frac{2m}{r_{rtp}}}},\frac{1+\frac{2m}{r_{tp}}}{1-\frac{2m}{r_{tp}}}\right],
\nonumber
\end{eqnarray}
where ${\rm EllipticF}[x,a]$ is an elliptic function of first kind. For the general case, we have to perform the integration numerically. Figure \ref{fig:sd} shows the plots for the deflection angle, obtained numerically for different values of the metric parameters. It should be noted that the deflection angle is negative for some parameter values. A negative deflection angle has also been obtained  
in gravitational lensing by a naked singularity \cite{naked1}. To visualize the negative deflection, we solve the geodesic equation and plot the light trajectories in Fig. \ref{fig:light_ray0}. The negatively deflected light rays are deflected away from the wormhole. Note that, for some parameter values, there is a turning point for which the net deflection angle becomes zero. We denote this turning point $r_{tp}$ by $r_{\Delta\phi=0}$. We name this as a {\em 
surface of zero deflection} in the sense that any light ray which comes from infinity, turns at the surface of radius $r_{\Delta\phi=0}$ and then goes to infinity without taking any turn around the wormhole lens, does not undergo any 
net deflection by the background gravitational field. This scenario is achieved if the light ray undergoes both positive and negative deflection (so that the net is zero) along its path. Figure \ref{fig:light_ray00} shows the equatorial trajectory of a light ray undergoing zero net deflection.

From Fig. \ref{fig:sd}, it is clear that the deflection angle for all turning points always remain positive if the weak deflection is positive. Therefore, 
it is useful to obtain an exact expression for the deflection angle in the weak field limit and identify the parameter region where such negative deflection occurs. The radial component of the null geodesic equation can be written as
\begin{equation}
2B\ddot{r}+B'\dot{r}^2+E^2\frac{A'}{A^2}-\frac{L^2}{C^2}C'=0.
\nonumber
\end{equation}
Using the inverse radial coordinates $u=\frac{1}{r}$ and $\dot{\phi}=\frac{L}{C}$, the above equation can be written in the form \cite{weak_lensing}
\begin{equation}
\frac{d^2u}{d\phi^2}+\frac{C}{B}u^3=-\frac{1}{2}u^2\frac{d}{du}\left(\frac{C}{B}u^2\right)+\frac{1}{2b^2}\frac{d}{du}\left(\frac{C^2u^4}{AB}\right).
\end{equation}
In our case,
\begin{equation}
A(u)=\frac{\left[(1+\frac{\beta}{m^2}m u)+\eta \sqrt{1-2m u -\frac{\beta}{m^2} m^2u^2}\right]^2}{(1+\eta)^2},
\nonumber
\end{equation}
\begin{equation}
B(u)=(1-2m u -\frac{\beta}{m^2} m^2u^2)^{-1}, \hspace{0.5cm} C(u)=\frac{1}{u^2},
\nonumber
\end{equation}
where $\frac{\beta}{m^2}=\frac{1-\mu^2}{\mu^2}$. For small deflection, we solve the above $u$ equation perturbatively. For small $u$, we obtain
\begin{equation}
\frac{d^2u}{d\phi^2}+u\simeq 3mu^2+2\beta u^3+\frac{1}{b^2}(-p+qu),
\label{eq:u_eqn}
\end{equation}
where
\begin{equation}
p=\frac{m}{\mu^2(1+\eta)}, \hspace{0.2cm} q=\frac{3m^2(1-\mu^2-\mu^2\eta)}{\mu^4(1+\eta)^2}.
\nonumber
\end{equation}
We now define a smallness parameter $\epsilon=m u_N$, where $u_N=1/b$ is the inverse of the Newtonian distance of closest approach. Therefore, in terms of the dimensionless variable $\xi=u/u_N$, Eqn. (\ref{eq:u_eqn}) yields
\begin{equation}
\frac{d^2\xi}{d\phi^2}+\xi=3\epsilon\xi^2+\frac{2\beta}{m^2}\epsilon^2\xi^3-\frac{p}{m}\epsilon+\frac{q}{m^2}\epsilon^2\xi.
\end{equation}
Expanding $\xi$ as $\xi=\xi_0+\epsilon\xi_1+\epsilon^2\xi_2+....$, we obtain the following order by order equations
\begin{equation}
\frac{d^2\xi_0}{d\phi^2}+\xi_0=0,
\nonumber
\end{equation}
\begin{equation}
\frac{d^2\xi_1}{d\phi^2}+\xi_1=-\frac{p}{m}+3\xi_0^2,
\nonumber
\end{equation}
\begin{equation}
\frac{d^2\xi_2}{d\phi^2}+\xi_2=\frac{q}{m^2}\xi_0+\frac{2\beta}{m^2}\xi_0^3+6\xi_0\xi_1.
\nonumber
\end{equation}
Solving these equations, we finally arrive at
\begin{eqnarray}
\frac{u}{u_N} &\simeq & \cos\phi-\left[\frac{p}{m}-\frac{3}{2}+\frac{\cos(2\phi)}{2}\right]\frac{m}{b}-\left[\left(\frac{\beta}{2m^2}-\frac{3}{2}\right)\frac{\cos(3\phi)}{8}
\right. \nonumber \\
& & \left. +\left(\frac{3p}{m}-\frac{q}{2m^2}-\frac{3\beta}{4m^2}-\frac{15}{4}\right)\phi\sin\phi \right]\frac{m^2}{b^2},
\nonumber
\end{eqnarray}
where it is assumed that the light source is at $\phi=\frac{\pi}{2}$. For small deflection, we have $u=0$ and $\phi=\frac{\pi}{2}+\delta$ at spatial infinity. The deflection angle $\Delta\phi$ ($=2\delta$) is given by
\begin{equation}
\Delta\phi \simeq \left(1-\frac{p}{2m}\right)\frac{4m}{b}+\left(-\frac{3p}{m}+\frac{q}{2m^2}+\frac{3\beta}{4m^2}+\frac{15}{4}\right)\pi \frac{m^2}{b^2}.
\end{equation}
At the turning point $r_{tp}=\frac{1}{u(0)}$, $\phi=0$. This gives the following relation between $r_{tp}$ and $b$:
\begin{equation}
\frac{1}{b}\simeq \frac{1}{r_{tp}}+\left(\frac{p}{m}-1\right)\frac{m}{r_{tp}^2}.
\end{equation}
In terms of $r_{tp}$, the deflection angle becomes
\begin{eqnarray}
\Delta\phi &\simeq& \left(1-\frac{p}{2m}\right)\frac{4m}{r_{tp}}+\left(1-\frac{p}{2m}\right)\left(\frac{p}{m}-1\right)\frac{4m^2}{r_{tp}^2} \nonumber \\
& & +\left(-\frac{3p}{m}+\frac{q}{2m^2}+\frac{3\beta}{4m^2}+\frac{15}{4}\right)\pi \frac{m^2}{r_{tp}^2}.
\end{eqnarray}
In the limit $\eta \to \infty$, $m=M$ and $\beta=-Q^2$, we recover the Reissner-Nordstr$\ddot{\text{o}}$m spacetime. Therefore, the deflection angle, in this limit, becomes
\begin{equation}
\Delta\phi\big|_{RN} \simeq \frac{4M}{r_{tp}}+\left[\left(-\frac{3Q^2}{4M^2}+\frac{15}{4}\right)\pi-4\right]\frac{M^2}{r_{tp}^2},
\end{equation}
which is the same as that obtained in \cite{RNL}. In terms of the constants $\mu$ and $\eta$, the weak deflection angle can be written as
\begin{equation}
\Delta\phi \simeq \left[1-\frac{1}{2\mu^2(1+\eta)}\right]\frac{4m}{b}+\left[-\frac{1}{\mu^2(1+\eta)}+\frac{1-\mu^2(1+\eta)}{2\mu^4(1+\eta)^2}+\frac{1-\mu^2}{4\mu^2}+\frac{5}{4}\right]3\pi \frac{m^2}{b^2}.
\label{eq:weak_deflection}
\end{equation}
It is to be noted that, in addition to the weak lensing by Schwarzschild mass $m$ (ADM mass), the naked singularities have positive ($\eta< -1$) and 
wormhole solutions have negative ($\eta>-1$) contribution to the first-order weak deflection. The first-order weak deflection angle is positive for $\mu^2(1+\eta)>\frac{1}{2}$, i.e., $(q_1-q_2)^2<4q_1^2 q_2^2$. For parameters not respecting this inequality, the weak deflection angle becomes negative.
\begin{figure}[ht]
\centering
\includegraphics[scale=1.0]{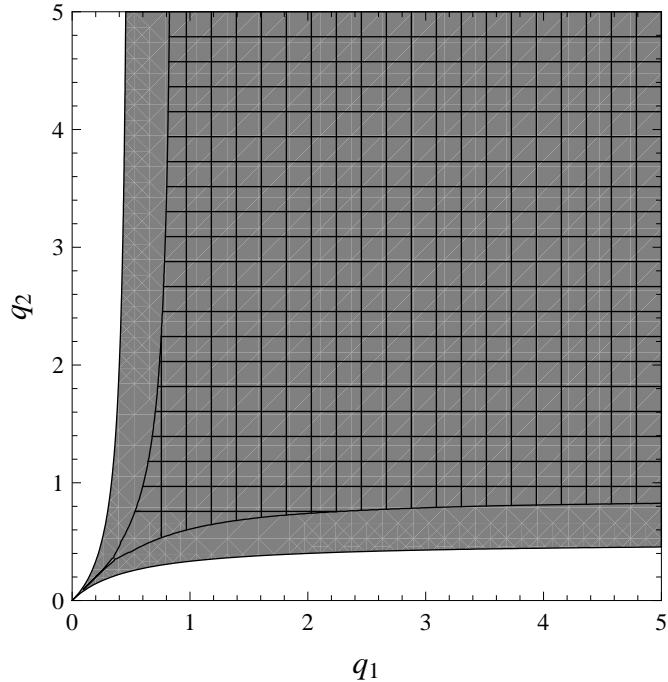}
\caption{Parameter regions showing the satisfaction of the energy conditions (box-shaded region) and the positivity of the weak deflection (gray-shaded region).}
\label{fig:wd}
\end{figure}

\begin{figure}[ht]
\centering
\includegraphics[scale=1.0]{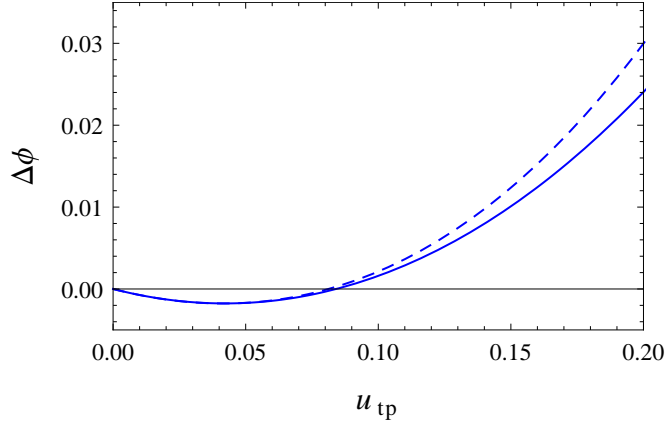}
\caption{Comparison between the numerically obtained exact deflection angle (solid curve) and
that found for weak deflection (dashed curve). Here, we have taken $\mu=0.8$, $\eta=-0.25$, $m=0.5$.}
\label{fig:zero}
\end{figure}
Figure \ref{fig:wd} shows the parameter regions for which the deflection angle is positive and the energy conditions are satisfied. It is clear that the deflection angle is always (i.e., for all impact parameters) positive for the parameter values satisfying the energy conditions. However, for the parameter values violating the energy conditions, the deflection angle can either be positive or negative depending on the impact parameter value. Figure \ref{fig:zero} shows a comparison between the numerically obtained deflection angle and that found for weak deflection. For $mu_{tp}\ll 1$, the matching of the two curves is reasonably
good, including the location of the turning point where zero deflection occurs. We mentioned earlier that the surface of zero deflection is achieved if the light ray undergoes both positive and negative deflection (so that the net is zero) along its path. Our analysis therefore indicates that the energy conditions must be violated to have a surface of zero deflection. 
Therefore, the existence of such a surface may be linked to the exotic matter 
(energy condition violating matter) content of the $R=0$ wormholes.
If matter satisfies the energy conditions, such a zero deflection surface 
will not exist.

\section{Gravitational lensing in the strong deflection limit}
\label{sec:strong deflection}
\begin{figure}[ht]
\centering
\includegraphics[scale=1.2]{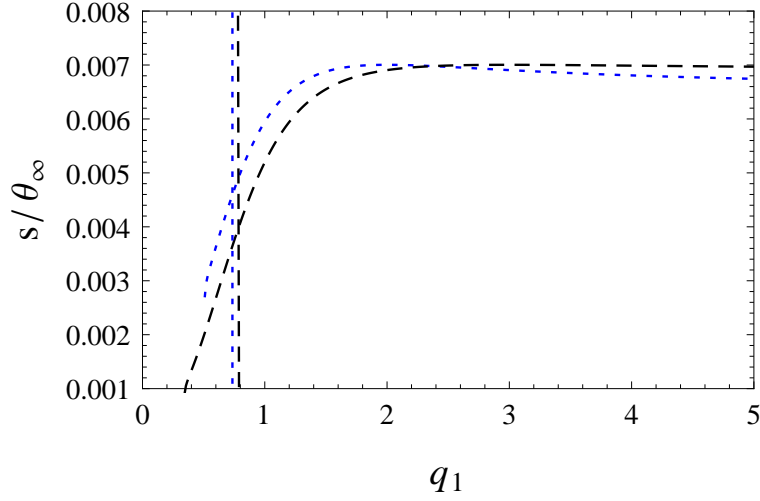}
\caption{Plot showing $\frac{s}{\theta_\infty}$ for $q_2=2$ (blue dotted) and $q_2=3$ (black dashed). The corresponding vertical lines show the separation between the parameter regions where the energy conditions are (right side) and are not (left side) satisfied.}
\label{fig:angular_diameter_1}
\end{figure}

\begin{figure}[ht]
\centering
\includegraphics[scale=1.2]{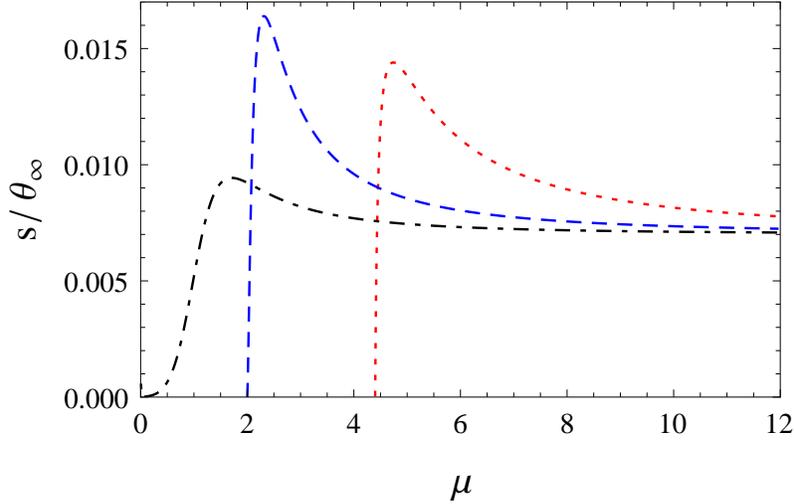}
\caption{Plot showing $\frac{s}{\theta_\infty}$ for $\eta=0.5$ (black dot-dashed), $\eta=0.0$ (blue dashed) and $\eta=-0.5$ (red dotted).}
\label{fig:angular_diameter_2}
\end{figure}

In the previous section, we have seen that the deflection angle may become negative. Therefore, it is useful to see what happens to the deflection angle in the strong deflection limit.  To study gravitational lensing in the strong deflection limit, we follow the analysis given by Bozza \cite{bozza_general1}. With the definition of two new variables
\begin{equation}
y=A(r), \hspace{0.5cm} v=\frac{y-y_{tp}}{1-y_{tp}},
\end{equation}
where $y_{tp}=A(r_{tp})$, the integral in the deflection angle becomes
\begin{equation}
I(r_{tp})=\int_0^1 R(v,r_{tp})F(v,r_{tp})dv,
\end{equation}
where
\begin{equation}
R(v,r_{tp})=2\frac{\sqrt{By}}{Cy'}(1-y_{tp})\sqrt{C_{tp}}
\end{equation}
and
\begin{equation}
F(v,r_{tp})=\frac{1}{\sqrt{y_{tp}-[(1-y_{tp})v+y_{tp}]C_{tp}/C}}.
\end{equation}
The term $R(v,r_{tp})$ is regular for all $v$ and $r_{tp}$. But, $F(v,r_{tp})$ diverges at $r=r_{tp}$, i.e., at $v=0$. In the limit $r\to r_{tp}$, $F(v,r_{tp})$ can be written as
\begin{equation}
F(v,r_{tp})\sim F_{tp}(v,r_{tp})=\frac{1}{\sqrt{\alpha v+\beta v^2}},
\end{equation}
where
\begin{equation}
\alpha=\frac{1-y_{tp}}{C_{tp}y'_{tp}}(C'_{tp}y_{tp}-C_{tp}y'_{tp}),
\end{equation}
\begin{equation}
\beta=\frac{(1-y_{tp})^2}{2C_{tp}^2y'^3_{tp}}[2C_{tp}C'_{tp}y'^2_{tp}+(C_{tp}C''_{tp}-2C'^2_{tp})y_{tp}y'_{tp}-C_{tp}C'_{tp}y_{tp}y''_{tp}].
\end{equation}
For the general metric (\ref{eq:arbitrary_metric}), the photon sphere radius $r_{ph}$ is given by $\frac{A'(r_{ph})}{A(r_{ph})}=\frac{C'(r_{ph})}{C(r_{ph})}$. Note that $\alpha$ vanishes at the photon sphere. Therefore, as the turning point $r_{tp}$ approaches the photon sphere $r_{ph}$, $I(r_{tp})$ diverges logarithmically. To show this, we split $I(r_{tp})$ into a divergent part $I_D(r_{tp})$ and a regular part $I_R(t_{tp})$ as
\begin{equation}
I(r_{tp})=I_D(r_{tp})+I_R(r_{tp}),
\end{equation}
where
\begin{equation}
I_D(r_{tp})=\int_0^1 R(0,r_{ph})F_{tp}(v,r_{tp})dv
\end{equation}
and
\begin{equation}
I_R(r_{tp})=\int_0^1 [R(v,r_{tp})F(v,r_{tp})-R(0,r_{ph})F_{tp}(v,r_{tp})]dv.
\end{equation}
After following the steps of \cite{bozza_general1}, the deflection angle in the strong deflection limit can be written as
\begin{equation}
\Delta\phi=-a\log\left(\frac{r_{tp}}{r_{ph}}-1\right)-\pi +b_D+b_R+\mathcal{O}(r_{tp}-r_{ph}),
\end{equation}
where
\begin{equation}
a=\frac{R(0,r_{ph})}{\sqrt{\beta_{ph}}}, \hspace{0.3cm} b_D=\frac{R(0,r_{ph})}{\sqrt{\beta_{ph}}}\log\left[\frac{2(1-y_{ph})}{y'_{ph}r_{ph}}\right],
\end{equation}
\begin{equation}
b_R=I_R({r_{ph}}),
\end{equation}
\begin{equation}
\beta_{ph}=\beta\big\vert_{r_{tp}=r_{ph}}=\frac{C_{ph}(1-y_{ph})^2}{2y^2_{ph}C'^2_{ph}}[C''_{ph}y_{ph}-C_{ph}y''_{ph}].
\end{equation}
In terms of the angular diameter $\theta$ of the relativistic images, the deflection angle can be written as
\begin{equation}
\Delta\phi=-\bar{a}\log\left(\frac{\theta D_{l}}{b_{ph}}-1\right)+\bar{b},
\end{equation}
where
\begin{equation}
\bar{a}=\frac{a}{2}, \hspace{0.3cm} \bar{b}=-\pi+b_R+\bar{a}\log\left(\frac{2\beta_{ph}}{y_{ph}}\right).
\end{equation}
Here, $D_l$ is the distance between the observer and the lens (see Fig. \ref{fig:schematic_lensing}). $b_{ph}$ is the minimum impact parameter given by $b_{ph}=\sqrt{\frac{C_{ph}}{y_{ph}}}$. For the light rays passing very close to the photon sphere $\Delta\phi\simeq 2\pi n$, where $n$ is the winding number. Therefore, the angular diameter of the $n{\rm th}$ relativistic image is given by
\begin{equation}
\theta_n=\theta_{\infty}\left[1+e^{\frac{\bar{b}-2\pi n}{\bar{a}}}\right],
\end{equation}
where $\theta_{\infty}=\frac{b_{ph}}{D_l}$. One of the observables is the angular separation between the outermost relativistic image ($\theta_1$) and the inner ones \cite{bozza_general1}. We have
\begin{equation}
s=\theta_1-\theta_{\infty}=\theta_{\infty}e^{\frac{\bar{b}-2\pi}{\bar{a}}}.
\end{equation}
Using Eqns. (\ref{eq:R0}) and (\ref{eq:extremum1}), we obtain, for the metric (\ref{eq:metric1}),
\begin{equation}
\bar{a}=\frac{1}{\sqrt{2-\frac{3m}{r_{ph}}}}, \hspace{0.3cm} \beta_{ph}=\frac{(1-f^2_{ph})^2}{4f^2_{ph}}\frac{\left(2-\frac{3m}{r_{ph}}\right)}{1-\frac{2m}{r_{ph}}-\frac{1-\mu^2}{\mu^2}\frac{m^2}{r^2_{ph}}},
\end{equation}
where $f_{ph}=f(r_{ph})$ and $r_{ph}=\frac{m(1+\mu)}{\mu z_{ex}}$, $z_{ex}$ being the solution of Eqn. (\ref{eq:extremum2}). We find $b_R$ numerically and plot $\frac{s}{\theta_{\infty}}$ in Fig. \ref{fig:angular_diameter_1} for real $q_1$ and $q_2$. Note that the relativistic images for the parameter values 
satisfying the energy conditions are less closely spaced than those for the 
parameter values violating the energy conditions. Figure \ref{fig:angular_diameter_2} shows the plot of $\frac{s}{\theta_\infty}$ for arbitrarily chosen $\mu$ and $\eta$ for which $q_1$ and $q_2$ may become complex conjugate to each other.

It should be noted that, in the limit $\mu\to\infty$, $\frac{s}{\theta_\infty}$ approaches a constant value. This can also be shown analytically. In the limit $\mu\to\infty$ (i.e., $\beta\to -m^2$), the metric becomes the extremal Reissner-Nordstr$\ddot{\text{o}}$m spacetime given by
\begin{equation}
ds^2=-\left(1-\frac{m}{r}\right)^2dt^2+\frac{dr^2}{\left(1-\frac{m}{r}\right)^2}+r^2d\Omega^2.
\end{equation}
For this metric, we find that $r_{ph}=2m$, $\beta_{ph}=\frac{9}{8}$ and
\begin{equation}
b_R=\int_0^1\left[\frac{6}{\sqrt{1+3v}}\frac{1}{\sqrt{1-(1+3v)(2-\sqrt{1+3v})^2}}-\frac{2\sqrt{2}}{v}\right]dv.
\end{equation}
The above integration can be done analytically if we put $w=(\sqrt{1+3v}-1)$. After performing the integration, we find that
\begin{equation}
b_R=2\sqrt{2}\log\left[\frac{4}{3}(2-\sqrt{2})\right].
\end{equation}
Therefore, we find that $\frac{s}{\theta_{\infty}}\simeq 0.007$, which matches with that in Fig. \ref{fig:angular_diameter_2}.

\section{Einstein and relativistic Einstein rings}
\label{sec:Einstein rings}

\begin{figure}[ht]
\centering
\includegraphics[scale=0.90]{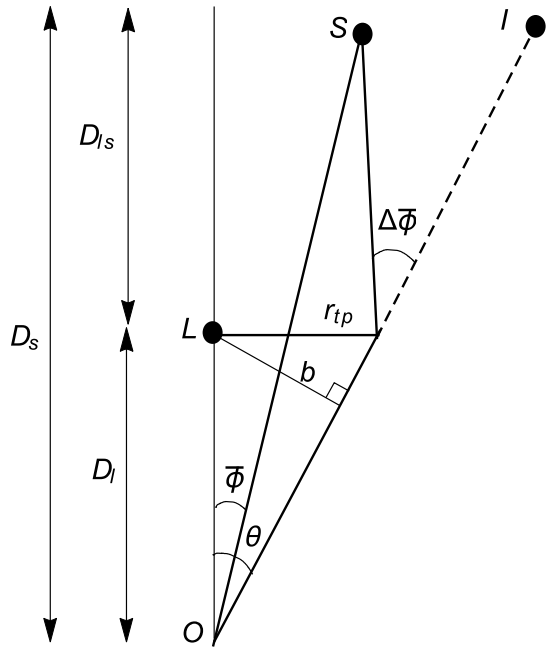}
\caption{Schematic diagram showing the configuration of gravitational lensing. The positions of the observer, the lens, the light source and the image are denoted by $O$, $L$, $S$ and $I$, respectively. $D_l$, $D_s$ and $D_{ls}$ are, respectively, observer-lens, observer-source and lens-source distance. $\Delta\bar{\phi}$ and $b$ are, respectively, the effective deflection angle and the impact parameter. $r_{tp}$ is the turning point and $\theta$ is the image position.}
\label{fig:schematic_lensing}
\end{figure}

In this section, we find the angular diameter of the Einstein and relativistic Einstein rings. The lensing configuration is shown in Fig. \ref{fig:schematic_lensing}.
When the observer and the source are far away from the lens, i.e., when $D_l\gg b$ and $D_{ls}\gg b$ (thin lens approximation), the lens equation becomes
\begin{equation}
D_{ls}\Delta\bar{\phi}=D_s(\theta-\bar{\phi}),
\end{equation}
where, $\theta=\frac{b}{D_l}$. Note that we have assumed $|\Delta\bar{\phi}|\ll 1$, $|\bar{\phi}|\ll 1$ and $|\theta|\ll 1$. For a ring image, we set $\bar{\phi}=0$. As the turning point approaches the photon sphere, the light rays wind around the lens many times. In general, the deflection angle $\Delta\phi$ can be expressed as $\Delta\phi=\Delta\bar{\phi}+2\pi n$. In the weak field limit ($b\gg r_0$), the winding number $n$ should be zero, thereby giving the Einstein ring. Therefore, the angular diameter of the Einstein ring is given by
\begin{equation}
\theta_0=\frac{D_{ls}}{D_s}\Delta\bar{\phi}=\frac{D_{ls}}{D_s}\Delta\phi=\frac{b}{D_l},
\label{eq:angular1}
\end{equation}
where $\Delta\phi$ is the weak deflection given in Eqn. (\ref{eq:weak_deflection}). Relativistic Einstein rings are formed by the light rays passing very close to the photon sphere and there are a large number (theoretically, an infinite number) of such rings. In the previous section, we have seen that these relativistic images are closely spaced. Here, we consider 
relativistic rings forming a single ring image \cite{virbhadra_schwarzschild1,harada1}. For such rings, the relation $\Delta\phi=\Delta\bar{\phi}+2\pi n$ is satisfied for $n\geq 1$ \cite{harada1} (since $\Delta\bar{\phi}$ is small). We may take $r_{tp}\simeq r_{ph}$, and the critical impact parameter is given by $b_{ph}= \sqrt{\frac{C_{ph}}{y_{ph}}}$. Therefore, the angular diameters of the relativistic Einstein rings become
\begin{equation}
\theta_{n\geq 1}=\frac{b_{ph}}{D_l}\simeq \frac{1}{D_l}\sqrt{\frac{C_{ph}}{y_{ph}}}.
\end{equation}
We then find out the expressions for the angular diameter $\theta_0$ and $\theta_{n\geq 1}$ and establish a relationship between them. Note that, in order to obtain the diameter angles and the relationship between them, we need an analytic expression for the photon sphere radius which is difficult to obtain for the general wormhole geometries. Therefore, we concentrate on the special case $\mu=1$. In this case, the first-order weak deflection angle is given by
\begin{equation}
\Delta\phi\simeq\left[1-\frac{1}{2(1+\eta)}\right]\frac{4m}{b}.
\end{equation}
The first-order weak deflection is negative for $-1<\eta<-\frac{1}{2}$. For $\eta >-\frac{1}{2}$, the angular diameter of the Einstein ring is given by (see Eqn. (\ref{eq:angular1}))
\begin{equation}
b\simeq\sqrt{\frac{2D_l D_{ls}}{D_s}\frac{2\eta +1}{1+\eta}m} \hspace{0.1cm} \Rightarrow \hspace{0.1cm} \theta_0=\frac{b}{D_l}\simeq \sqrt{\frac{2D_{ls}}{D_l D_s}\frac{2\eta +1}{1+\eta}m}.
\label{eq:Einstein_ring1}
\end{equation}
However, the first-order weak deflection is zero for $\eta=-\frac{1}{2}$. In this case, we have to consider the second-order term which is $\Delta\phi\simeq \frac{3\pi}{4}\frac{m^2}{b^2}$. Therefore, the angular diameter is given by
\begin{equation}
b\simeq \left(\frac{3\pi}{4}\frac{D_l D_{ls}}{D_s}m^2 \right)^{\frac{1}{3}} \hspace{0.1cm} \Rightarrow \hspace{0.1cm} \theta_0=\frac{b}{D_l}\simeq \left(\frac{3\pi}{4}\frac{D_{ls}}{D_l^2 D_s}m^2 \right)^{\frac{1}{3}}.
\label{eq:Einstein_ring2}
\end{equation}
To find the angular diameter of the relativistic Einstein rings, we need to find the photon sphere radius by solving Eqn. (\ref{eq:extremum2}) 
for $\mu=1$. For $\eta>0$, we obtain
\begin{equation}
r_{ph}=\frac{9m\eta^2}{3\eta^2-1+\sqrt{3\eta^2+1}}.
\end{equation}
The photon sphere does not exist for $-1<\eta\leq 0$. We recover the photon sphere radius $r_{ph}=3m$ of the Schwarzschild black hole for $\eta\to\infty$. It tends to $r_0=2m$ as $\eta\to 0$. After some manipulation, we obtain the critical impact parameter
\begin{equation}
b_{ph}=\frac{27m\left(1+\eta\right)\eta^2}{\left(3\eta^2+1\right)^{3/2}+\left(9\eta^2-1\right)},
\end{equation}
for light rays winding around the photon sphere. For large $\eta$, we recover the critical impact parameter $b_{ph}\simeq 3\sqrt{3}m$ of the Schwarzschild black hole. However, for $-1<\eta\leq 0$, the photon sphere does not exist and the deflection angle diverges as $r_{tp}$ approaches the wormhole throat $r_0$. In that case, we take $r_{tp}\simeq r_0= 2m$ to calculate the critical impact parameter which becomes $b_{ph}=2m\left(1+\eta\right)$. Therefore, the angular diameters of the relativistic Einstein rings become
\begin{equation}
\theta_{n\geq 1} \simeq \frac{b_{ph}}{D_l}=\left\{
  \begin{array}{ll}
  \frac{2m(1+\eta)}{D_l} & : -1<\eta\leq0\\
  \frac{27m\left(1+\eta\right)\eta^2}{\left(3\eta^2+1\right)^{3/2}+\left(9\eta^2-1\right)} \frac{1}{D_l} & : \eta> 0.
  \end{array}
\right.
\label{eq:relativistic_rings}
\end{equation}
The relationship between $\theta_{n\geq 1}$ and $\theta_0$ turns out to be
\begin{equation}
\theta_{n\geq 1} \simeq \left\{
  \begin{array}{ll}
  \sqrt{\frac{4}{3\pi}\frac{D_s}{D_{ls}}}\theta_0^{\frac{3}{2}} & : \eta=-\frac{1}{2}\\
  \frac{D_s}{D_{ls}}\frac{(1+\eta)^2}{2\eta+1}\theta_0^2 & : -\frac{1}{2}<\eta\leq 0\\
  \frac{\left(1+\eta\right)\eta^2}{\left(3\eta^2+1\right)^{3/2}+\left(9\eta^2-1\right)} \frac{1+\eta}{2\eta+1} \frac{27D_s}{2D_{ls}}\theta_0^2 & : \eta> 0.
  \end{array}
\right.
\nonumber
\end{equation}
Thus, the qualitative feature of lensing for $\eta=-\frac{1}{2}$ is different from that for $\eta>-\frac{1}{2}$.

\section{Detectability of rings and wormhole size}
\label{sec:detectability}

\noindent The authors in \cite{harada1} have shown that one can detect the relativistic 
Einstein rings formed by an Ellis-Bronnikov wormhole (i) with throat radius $0.5$ pc at a Galactic center and  
distance $D_l=D_{ls}=10$ Mpc and (ii) with throat radius 10 AU in 
our Galaxy and distance $D_l=D_{ls}=10$ kpc,
using the most powerful modern instruments which have resolution 
of $10^{-2}$ arcsecond (such as a 10-meter optical-infrared telescope). 
In order to detect the relativistic Einstein rings by a telescope with 
resolution of $10^{-2}$ arcsecond, the minimum angular diameter of the 
relativistic rings should be of the order of $10^{-2}$ arcsecond. 

\noindent What is the scenario of a similar detection possibility of these 
rings which may be formed by the type of wormholes discussed in this article?
Notice that for our wormholes, we have a parameter $\eta$ as well as $m$.
Using Eqn. (\ref{eq:relativistic_rings}), we find that, for a wormhole (of the
class discussed here) 
at a Galactic center with $D_l=D_{ls}=10$ Mpc, the minimum 
throat size ($r_0=2m$) should 
lie between 0.97 pc and 0.34 pc for $-0.5\leq\eta\leq0.5$. One 
can calculate the corresponding mass $M$ of the wormhole using $m=\frac{GM}{c^2}$. This turns out to be in the range between $1.01\times 10^{13} M_\odot$ and $3.57\times 10^{12} M_\odot$ for $-0.5\leq\eta\leq0.5$, $M_\odot$ being the mass of the Sun. 
However, to detect the relativistic Einstein rings produced by such 
a wormhole at the center of our Galaxy with $D_l=D_{ls}\simeq 10$ kpc, the 
throat size should lie between $200.01$ AU and $70.43$ AU for $-0.5\leq\eta\leq0.5$. 
The corresponding mass of the wormhole turns out to be in the range between $1.01\times 10^{10} M_\odot$ and $3.57\times 10^9 M_\odot$ for $-0.5\leq\eta\leq0.5$. The above results for these wormholes are therefore similar, in terms of
mass, to those for supermassive black holes \cite{harada1}.

\begin{figure}[ht]
\centering
\includegraphics[scale=1.2]{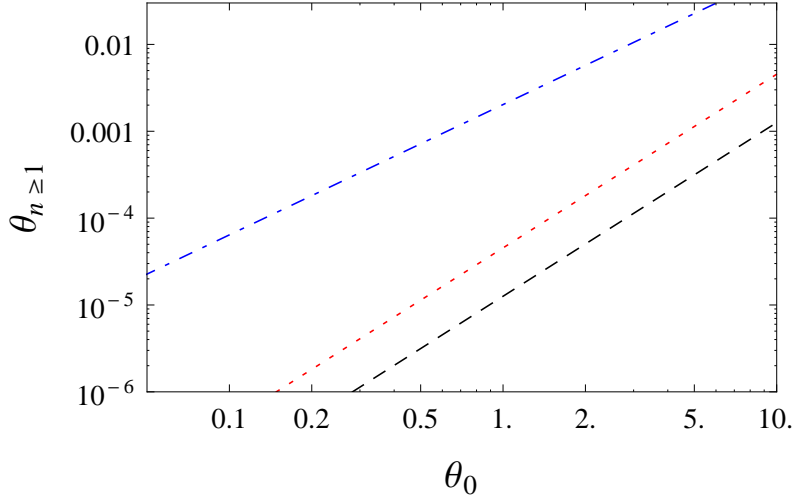}
\caption{LogLog plots showing (in arcseconds) the angle of the relativistic Einstein ring $\theta_{n\geq 1}$ versus the angle of the Einstein ring $\theta_0$ for $\eta=-0.5$ (blue dot-dashed), $\eta=-0.47$ (red dotted) and $\eta\to \infty$ (black dashed). Here, we have taken $D_l=D_{ls}=10$ Mpc. The case $\eta\to\infty$ represents black hole. In generating a plot for a given $\eta$, $m$ is used as a parameter.}
\label{fig:comparision}
\end{figure}

\noindent It may be noted that one can find the wormhole parameters 
$m$ and $\eta$ once $D_l$, $D_{ls}$, $\theta_0$ and $\theta_{n\geq 1}$ are given. Figure \ref{fig:comparision} shows (in arcseconds) the angle of the relativistic Einstein rings $\theta_{n\geq 1}$ versus the angle of the Einstein ring $\theta_0$ for different values of $\eta$. It is clear that from such a plot
it is possible to distinguish quantitatively  between a black hole ($\eta \rightarrow \infty$,
black-dashed line in Fig. 11) and wormholes ($\eta$ finite, the blue and red
lines in Fig. 11). 

\section{Conclusion}
\label{sec:conclusion}
After recalling the $R=0$ on-brane scalar-tensor wormhole solutions, we have briefly studied the energy conditions and have identified the parameter region that satisfies all the energy conditions. We first looked at the existence of  
stable and unstable photon orbits and its relation with the energy conditions, in the context of the scalar-tensor wormholes. 
We found that the energy conditions must be violated to have stable photon 
orbits. However, unstable photon orbits exist for parameter values satisfying or 
violating the energy conditions. Next, we have investigated 
gravitational lensing by this class of  wormholes and have tried to find a possible connection between lensing and the energy conditions. 
We find that the deflection of light is always positive, i.e., light rays always bend towards the wormholes if the energy conditions are satisfied. 
However, if the energy conditions are violated, light rays either bend towards or away from the wormholes depending on the parameter values. For some parameter values violating the energy conditions, we have found a turning point 
for which there is no net deflection even though we have a 
curved background geometry. We named this turning point as the surface of 
zero deflection. We have argued that the existence of such a surface is 
linked to the exotic matter content of our wormholes. Finally, we
briefly explored gravitational lensing in the strong field limit and obtain 
the fractional angular separation between the outermost relativistic image and the inner ones. It is found that the relativistic images for the parameter 
values satisfying the energy conditions, are less closely spaced than those 
for parameter values violating the energy conditions. We also
analyzed the formation of Einstein and relativistic Einstein rings and 
have obtained expressions for their diameter angles in terms of the metric 
parameters. The detectability of these rings as well as estimates on the sizes and masses of the wormholes have been addressed.

The above results clearly suggest a link between
wormhole existence, energy conditions and lensing features. 
Our conclusions are largely based on the line element we have chosen
to work with. However, we do believe that most of our results, in some general
form, do carry over to a broader class of wormholes in different alternative theories of gravity. We hope to 
arrive at more general statements in our future investigations. As further investigations, we would like to study gravitational microlensing (light curves) and retrolensing by these wormholes and find possible connections with the energy conditions.

\section*{Acknowledgment}
\noindent R. S. acknowledges the Council of Scientific and Industrial Research, Government of India for providing support through a fellowship.

\end{document}